# Thermal rectification in jointless Pb solid wire

Masayuki Mashiko,[1,†] Poonam Rani,[1,†], Yuto Watanabe[1], and Yoshikazu Mizuguchi[1]*

[1]*Department of Physics, Tokyo Metropolitan University, 1-1, Minami-osawa, Hachioji, 192-0397, Japan*

(† These authors contributed equally.)

* mizugu@tmu.ac.jp

**Abstract:**

Thermal rectification is observed in jointless Pb wires at temperatures near the superconducting transition of Pb under magnetic fields. Using different magnetic-field ($H$) response of temperature dependence of thermal conductivity ($\kappa$-$T$) under $H // J$ and $H \perp J$ where $J$ is heat flow, we fabricated a jointless thermal diode. Thermal rectification is observed with the thermal rectification ratio (TRR) of 1.5 and the difference in $\kappa$ of 330 W m$^{-1}$ K$^{-1}$ at $T$ = 5.11 K under $H$ = 400 Oe for a Pb wire with a 50%-bent ($H \perp J$) and 50%-straight ($H // J$) structure. The peak temperature of TRR can be tuned by the strength of applied magnetic field. By changing bent ratio to 40%-bent, a higher TRR exceeding 2 was observed. The Pb-jointless thermal diode will be a useful material for thermal management at cryogenic temperatures.

**Impact statement:**

Thermal rectification was observed in jointless solid wires of Pb by bending the half of the wire to expose magnetic fields parallel and perpendicular to the heat direction, $H // J$ and $H \perp J$. Because of the different thermal-conductivity response to the field direction ($H // J$ and $H \perp J$), the half-bend Pb wire works as thermal diode. Development of diode without joint will provide an impact strategy for diode design.

**Introduction**

Superconductivity is a quantum phenomenon characterized by the zero electrical resistivity and the exclusion of magnetic flux at temperatures below a superconducting transition temperature ($T_c$). The zero-resistivity states have been used in high-field magnet application and superconducting power cables for large-scale electricity transport. In addition, Josephson junctions comprising superconductors separated by a non-superconducting thin layer have been used in numerous sensors and quantum computers.[1–5] Furthermore, in the superconducting states, electronic thermal conductivity ($\kappa_{el}$) is suppressed due to the formation of Cooper pairs, and the low thermal conductivity ($\kappa$) states are used in current leads of superconducting magnets.[6] The suppression of $\kappa$ can be controlled by applied magnetic field ($H$), which achieves magneto-thermal switching (MTS).[7,8] In our recent paper,[9] we reported that the MTS ratio (MTSR), calculated as MTSR $(T, H) = [\kappa(T, H) - \kappa(T, H = 0\ \text{Oe})] / \kappa(T, H = 0\ \text{Oe})$ where $T$ and $H$ denote temperature and magnetic field, respectively, of high-purity (5N purity) Pb polycrystalline wire reached 2000% at $T = 2.5$ K.[9] Because MTS is one of the important technologies in the field of thermal management,[10-15] the use of superconductor MTS is expected to improve the efficiency of cryogenic devices. In addition, the notable feature of $\kappa$-$T$ for the Pb-5N wire is a sharp reduction of $\kappa$ at $T_c$ when the applied $H$ is parallel to the heat flow ($J$) direction.[16] As shown in Figs. 1(a) and 1(b), the $\kappa$-$T$ for the Pb-5N wire shows broad reduction of $\kappa$ when the applied $H$ is perpendicular to $J$. The difference motivated us to fabricate jointless thermal diode using the Pb-5N wires.

Diode is used in an electronic circuit for electricity rectification. The components of the diode are semiconductor with different carrier type. The thermal-transport counterpart of electric diode is called thermal diode.[12,17,18] The conventional thermal diode is composed of two materials having different thermal transport (value of $\kappa$ or $\kappa$-$T$ characteristics), and thermal rectification is observed under the temperature difference ($\Delta T$) between two edges of the diode. For example, in a bulk solid-state thermal diode made by $Ag_2(S,Se,Te)$, the thermal rectification efficiency was typically investigated under the $\Delta T$ of about 100 K.[18] For cryogenic temperature range, thermal diode based on superconductor-normal conductor junction was proposed in 2013.[19,20] The concept of superconductor-based thermal diodes (SC thermal diode) is based on the junction made of superconductor and normal metal or the use of Josephson junction, which is composed of superconductor-insulator (or normal metal)-superconductor junction. Most studies on SC thermal diodes have been based on simulation and investigation of nano-scale devices at low temperatures (lower than 1 K).[21–23]

Here, we designed a jointless thermal diode using Pb-5N wire after the idea that a junction composed of two materials with different thermal transport can work as a thermal diode.[19,20] In addition, if thermal rectification is achieved in a single wire without any joints, the flexibility of diode size and form will be merit in practical application in cryogenic devices. We use the difference in $\kappa$-$T$ between $H // J$ and $H \perp J$. To obtain the sample condition close to $H \perp J$ in the half of the sample, the Pb wire was bent (half-bent Pb wire) as shown in Fig. 1(c). The W-shaped part feels $H$ almost perpendicular to $J$. As shown in Fig. 1(d), the measurement directions, forward and reverse directions, are defined.

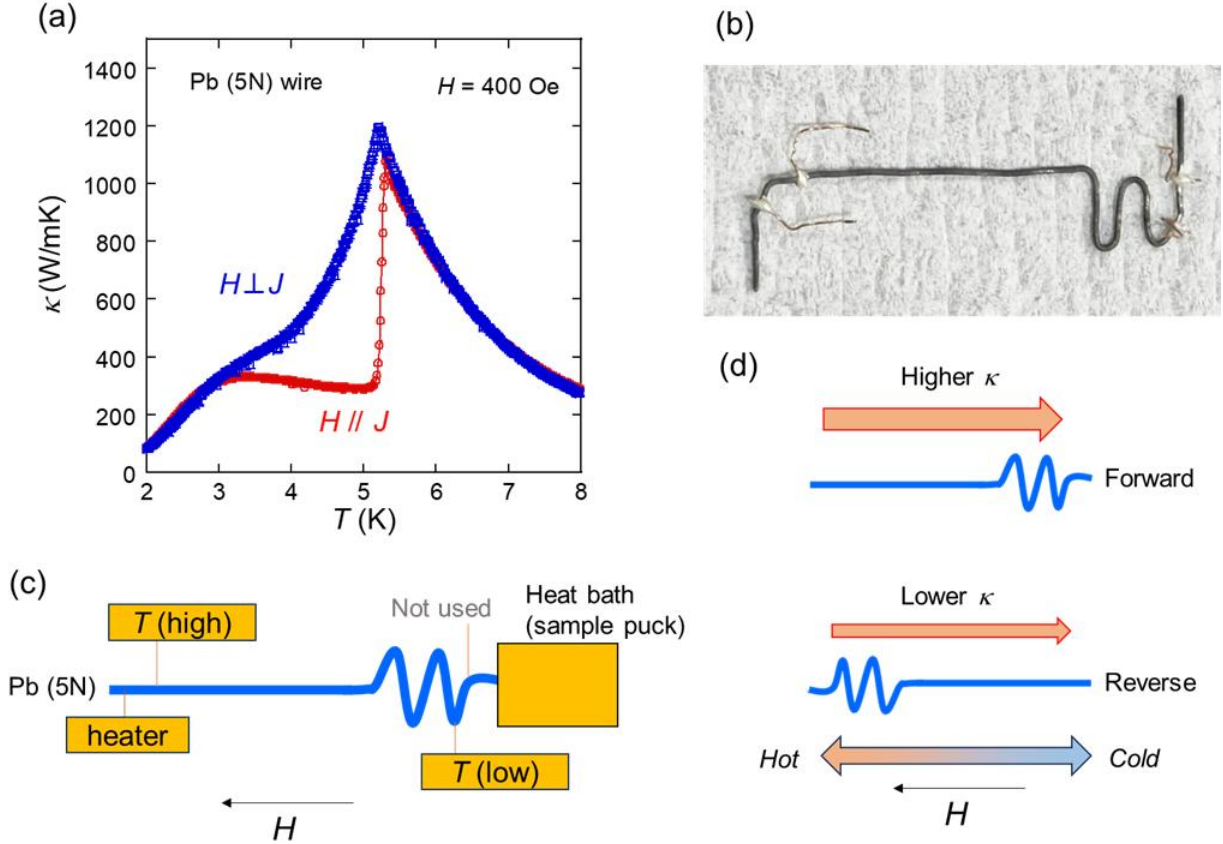

FIG. 1. (a) $\kappa$-$T$ of the Pb-5N wire under magnetic fields of $H // J$ and $H \perp J$ (400 Oe). (b) Photo of the prepared half-bent (50%-bent and 50%-straight) Pb-wire sample with electrodes (after measurements). The distance between two thermometers is 30.58 mm. (c) Schematic image of four terminal method for thermal conductivity measurement for the sample on TTO sample puck. (d) Definition of the heat flow directions, forward and reverse directions. The original data for $\kappa$-$T$ with $H // J$ has been reported in Ref. 16.

**Materials and methods**

The Pb (5N purity) polycrystalline wires with a diameter of 0.5 mm were purchased from the Nilaco Corporation. Measurements of $\kappa$ with four-terminal methods were performed on Physical Property Measurement System (PPMS-Dynacool, Quantum Design) using the thermal transport option (TTO). For the $\kappa$ measurements, the terminals were fabricated using Ag paste and Cu wires with a diameter of 0.2 mm, and the field direction ($H // J$) was controlled by changing the sample setup. The typical measurement period was 5-10 s, and the $T$ sweep speed for $\kappa$-$T$ was 0.05 K min$^{-1}$. To obtain relatively large temperature difference between two thermometers, the target temperature rise was fixed to 10%; we confirmed that this condition results in a good waveform in the $\kappa$ measurements. For the $\kappa$-$T$ measurements shown in Fig.1, the target temperature rise of 3% was used. In the plots of this paper, $T$ is average temperature of the sample under TTO measurements. Electrical resistivity of the Pb wire (pressed plate) was measured by four-terminal method with DC current of 3.5 mA on PPMS-Dynacool with electrodes made by Au wires and Ag paste. The field direction was controlled by a rotator probe.

**Results and discussion**

Figure 2 summarizes the thermal rectification properties of the half-bent (50%-bent and 50%-straight) Pb-5N wire (sample #1). Reproducibility has been confirmed by measuring three different samples, and selected data for sample #2 and #3 is summarized in Figs. S1 and S2. As shown in Fig. 2(a–e), the difference between $\kappa_F$-$T$ and $\kappa_R$-$T$ is clearly observed under magnetic fields while the data at $H$ = 0 Oe are almost the same. Noticeably, the temperature where the $\kappa$ drops due to superconducting transition is clearly different at $H$ = 200 and 400 Oe, which is caused by the different $\kappa$-$T$ characteristics with $H // J$ and $H \perp J$. Because of the difference between $\kappa_F$-$T$ and $\kappa_R$-$T$, $\Delta\kappa$-$T$ exhibits the peak structure under magnetic fields as shown in Figs. 2(f–j). TRR-$T$ is shown in Figs. 2(k–o); the highest TRR reaching 1.4 is observed at $T$ = 5 K and $H$ = 400 Oe. Through the $T$-sweep measurements of $\kappa$, thermal rectification was confirmed in the half-bent Pb wire. Although the absolute values of $\kappa_F$ and $\kappa_R$ are slightly different for the examined three samples, the overall trends, TRR, and the peak position of TRR are reproduced as shown in Figs. S1 and S2. For further confirmation, we measured $\kappa_F$ and $\kappa_R$ with similar $\Delta T$ because, by measuring $\kappa_F$ and $\kappa_R$, the ratio of heat current density ($J_F$ and $J_R$); the results obtained at $H$ = 400 Oe are summarized in Table I. The highest TRR of 1.53 is obtained at $T$ = 5.11 K.

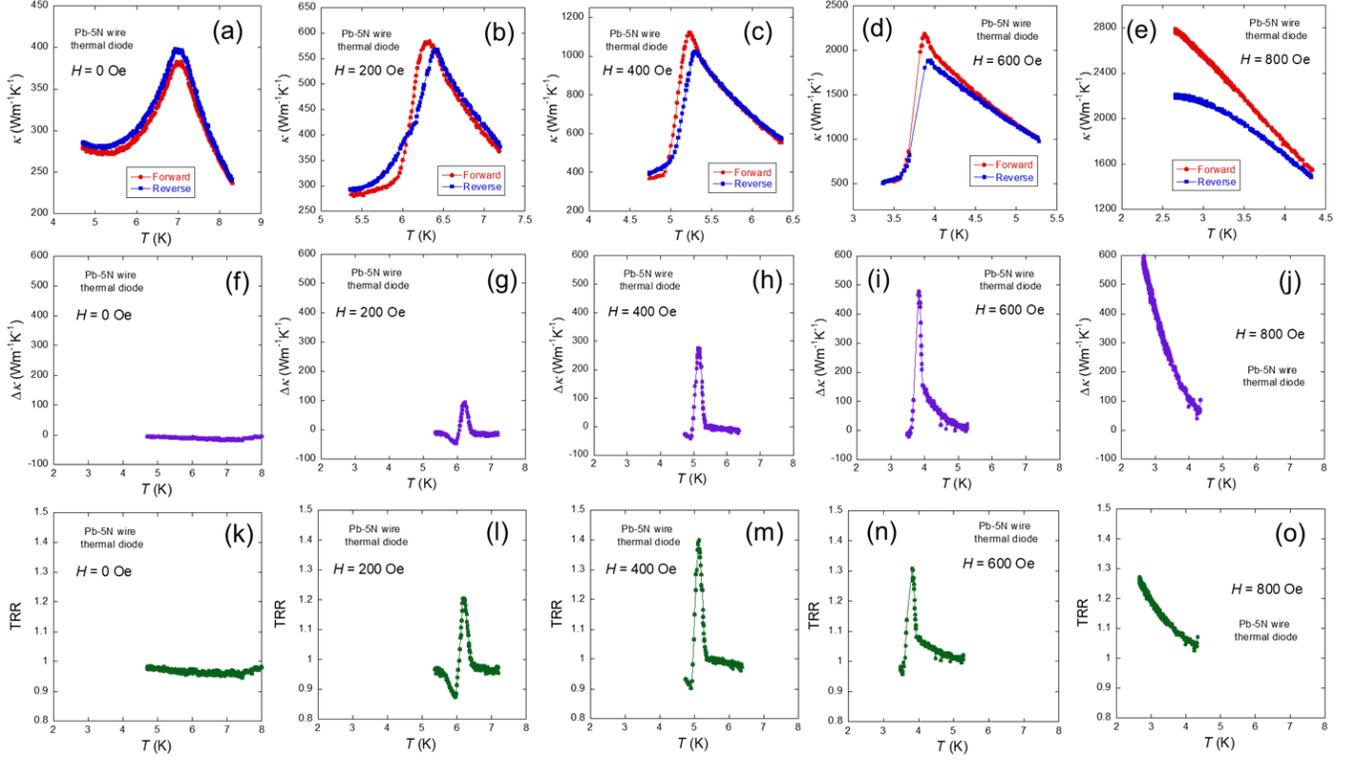

FIG. 2. (a–e) $\kappa$-$T$ of the bent Pb-5N wire (sample #1) measured along the forward and reverse directions at $H$ = (a) 0, (b) 200, (c) 400, (d) 600, and (e) 800 Oe. (f–j) $\Delta\kappa$-$T$ of the bent Pb-5N wire (sample #1) at $H$ = (f) 0, (g) 200, (h) 400, (i) 600, and (j) 800 Oe. (k–o) TRR-$T$ of the bent Pb-5N wire (sample #1) at $H$ = (k) 0, (l) 200, (m) 400, (n) 600, and (o) 800 Oe.

Table I. Thermal conductivity of the half-bent Pb wire measured at $H$ = 400 Oe with similar $\Delta T$.

| $T$ (K) | $\kappa_F$ (W m$^{-1}$ K$^{-1}$) | $\kappa_R$ (W m$^{-1}$ K$^{-1}$) | $\Delta T$ (K) | TRR |
|---|---|---|---|---|
| 5.23 | 1092 | 911 | 0.23 | 1.20 |
| 5.2 | 1092 | 851 | 0.25 | 1.28 |
| 5.17 | 1070 | 778 | 0.26 | 1.38 |
| 5.14 | 1015 | 705 | 0.27 | 1.44 |
| 5.11 | 947 | 617 | 0.29 | 1.53 |

Although we observed clear thermal rectification in the half-bent Pb wires, the mechanism for the emergence of different $\kappa$-$T$ with $H$ // $J$ and $H \perp J$ has not been addressed so far. First, we show the $\kappa$-$T$ data with various $H$ // $J$ and $H \perp J$ in Fig. 3. For all the fields, a sharp drop of $\kappa$ is seen under $H$ // $J$, and a broader transition is seen under $H \perp J$. This clarifies that the emerging phenomena are essentially driven by the applied-field direction. Figure 4 shows the temperature dependence of electrical resistivity ($\rho$-$T$) of the pressed Pb-5N wire measured at $H$ = 0, 200, 400, and 600 Oe with $H \mathbin{/\mkern-5mu/} I$ and $H \perp I$ where $I$ is electrical current, and the current is generated along the wire-length direction. At H = 200, 400, and 600 Oe, clear difference in the $\rho$-$T$ is seen between $H \mathbin{/\mkern-5mu/} I$ and $H \perp I$. For confirmation of reproducibility, different Pb-5N sample was investigated, and the $\rho$-$T$ at $H$ = 400 Oe is shown in Fig. S3. The zero-resistivity temperature is lower for $H \perp I$. Therefore, the difference in $\kappa$-$T$ with $H \mathbin{/\mkern-5mu/} J$ and $H \perp J$ is directly linked to the difference in superconducting properties of the Pb wire. Although the detailed mechanism for the different superconducting states is unclear, there should be difference in demagnetization effects due to the different size and shape to the magnetic field direction. In addition, as discussed in Refs. 24 and 25, intermediate states where normal-conducting regions are generated and accepting magnetic fluxes appear inside the Pb sample under magnetic field. In Ref. 24, formation of intermediate states and the suppression of superconducting current path with magnetic field perpendicular to the current in a rod sample was described. We assume that the uniform intermediate states are formed along the wire direction, and the superconducting regions are less affected by the applied field when $H \mathbin{/\mkern-5mu/} J$. The detailed clarification of the intermediate states should be addressed in a future work because the difference in the superconducting states in the half-bent Pb wire is essential for the emergence of thermal rectification.

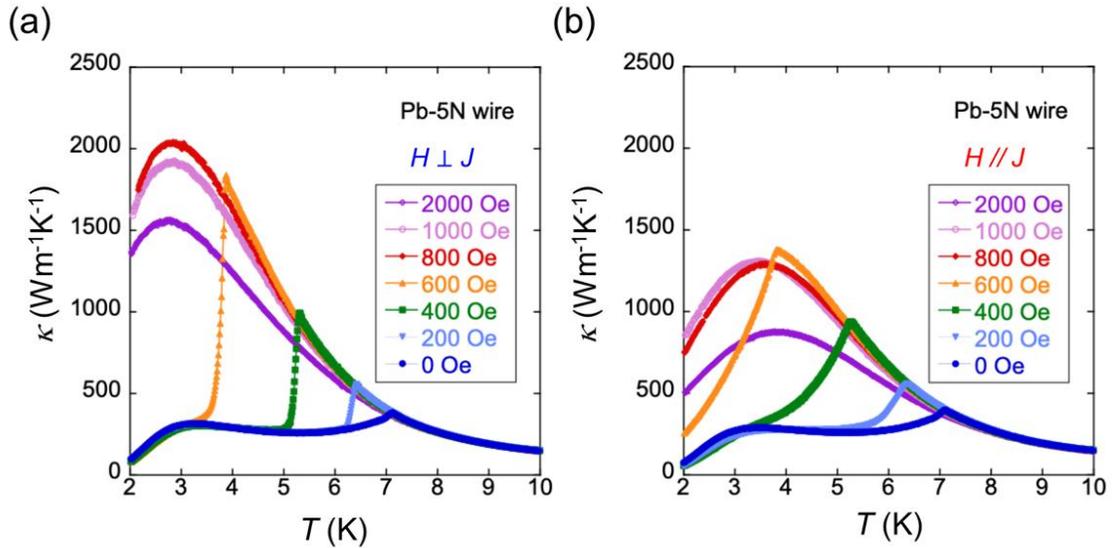

FIG. 3. Temperature dependence of $\kappa$ of 5N-Pb wire measured under magnetic field (a) parallel and (b) perpendicular to the wire length direction (heat flow direction).

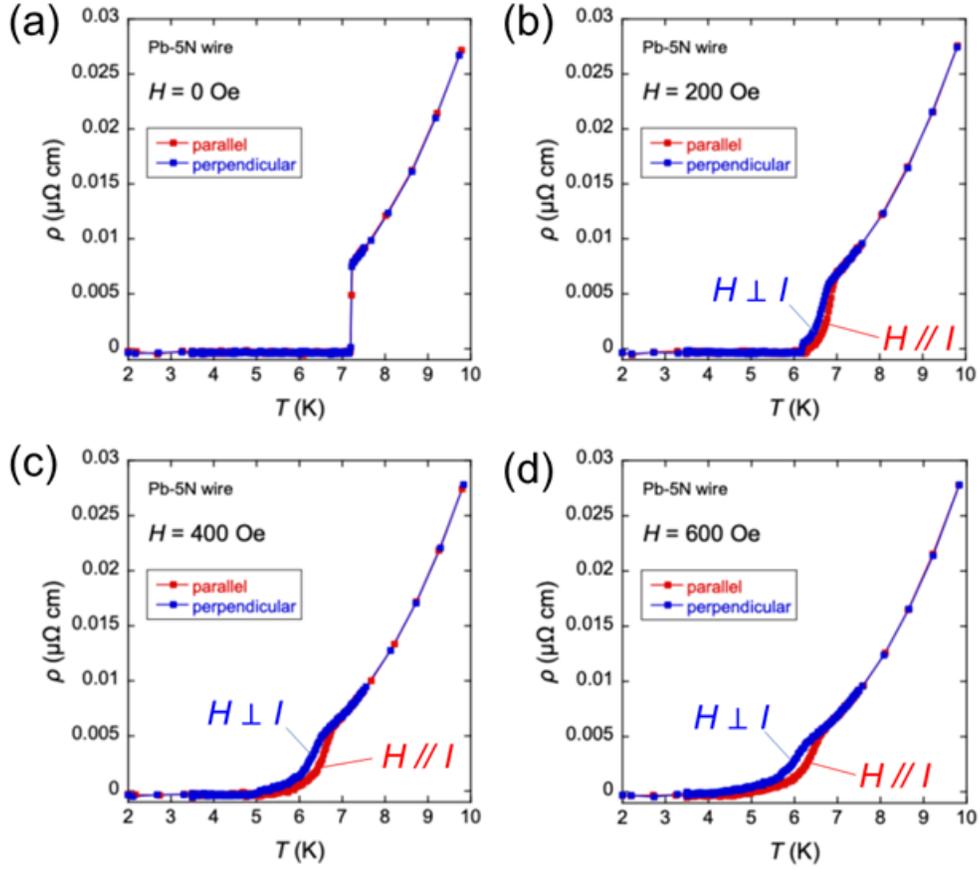

FIG 4. Temperature dependence of electrical resistivity ($\rho$-$T$) of the pressed Pb-5N wire measured at $H$ = 0, 200, 400, and 600 Oe with $\boldsymbol{H} \mathbin{/\mkern-4mu/} \boldsymbol{I}$ and $\boldsymbol{H} \perp \boldsymbol{I}$.

Next, we briefly investigate the effect of bent ratio on the thermal rectification properties. The photo of the prepared samples is displayed in Fig. S4. In Fig. S5 and S6, the results of the thermal transport measurements for the 40%-bent and 60%-bent samples are summarized. Due to the structural limitation, we currently succeeded in the fabrication and the TRR measurements for 40–60% ratio samples. Similar thermal rectification was observed for all the samples. The estimated maximum TRR at the peak field is plotted in Fig. 5 as a function of $H$. For the 60%-bent and 50%-bent samples, similar $H$ dependence of TRR is observed. For the 40%-bent sample, a higher TRR is observed, and the TRR tends to increase with increasing $H$; the highest TRR is 2.4 at $H$ = 600 Oe. The results suggest that the thermal rectification performance of the bent Pb-5N wire can be further improved by optimizing the diode structure.

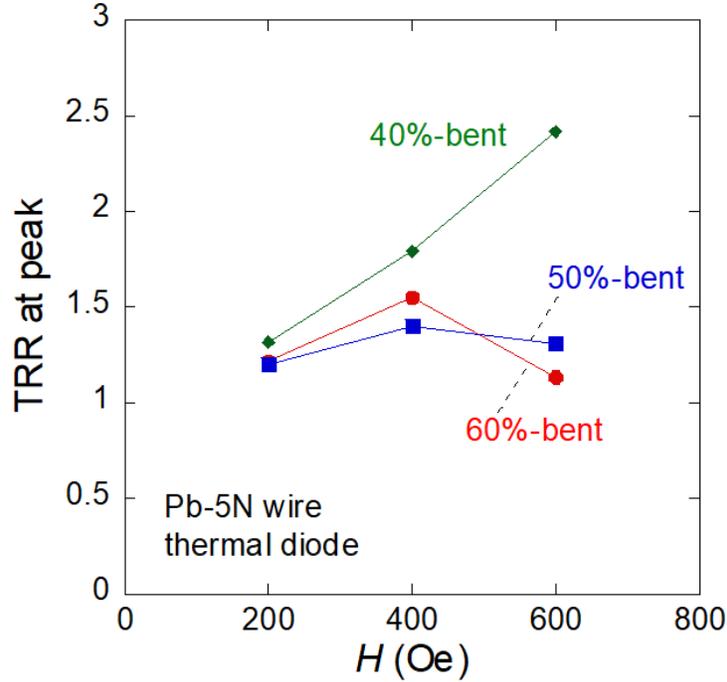

Fig. 5. Magnetic-field dependence of the maximum TRR at the peak field for 40%-, 50%-, and 60%-bent Pb-5N wire thermal diode.

In conclusion, we fabricated half-bent Pb wire by bending the half of Pb-5N wire to create both regions that feel $H \parallel J$ or $H \perp J$ in a single wire sample. Using the different $\kappa$-$T$ characteristics with $H \parallel J$ and $H \perp J$, thermal rectification was achieved. TRR of 1.53 and $\Delta\kappa$ of 330 W m$^{-1}$ K$^{-1}$ were obtained at $T = 5.11$ K and $H = 400$ Oe for the 50%-bent sample. A higher TRR exceeding 2 was observed for a 40%-bent sample. We expect that similar thermal diode can be developed using high-purity-metal superconductor like Sn, In, and Ta. The observation of thermal rectification in jointless bulk superconductor wire will provide us with new strategy for efficient thermal management in cryogenic devices.

**Acknowledgements**

The authors thank H. Usui, K. Uchida, and K. Hirata for valuable discussions. The work was partly supported by JST-ERATO (JPMJER2201) and TMU research fund for young scientists.

**Declaration of interest statements**

The authors declare no competing interests.

**SUPPLEMENTARY MATERIALS**

# Thermal rectification in jointless Pb solid wire

**Masayuki Mashiko[1,†], Poonam Rani[1,†], Yuto Watanabe[1], and Yoshikazu Mizuguchi[1,*]**

[1] Department of Physics, Tokyo Metropolitan University, Hachioji, Japan

[*]E-mail: mizugu@tmu.ac.jp
[†] Equal contribution

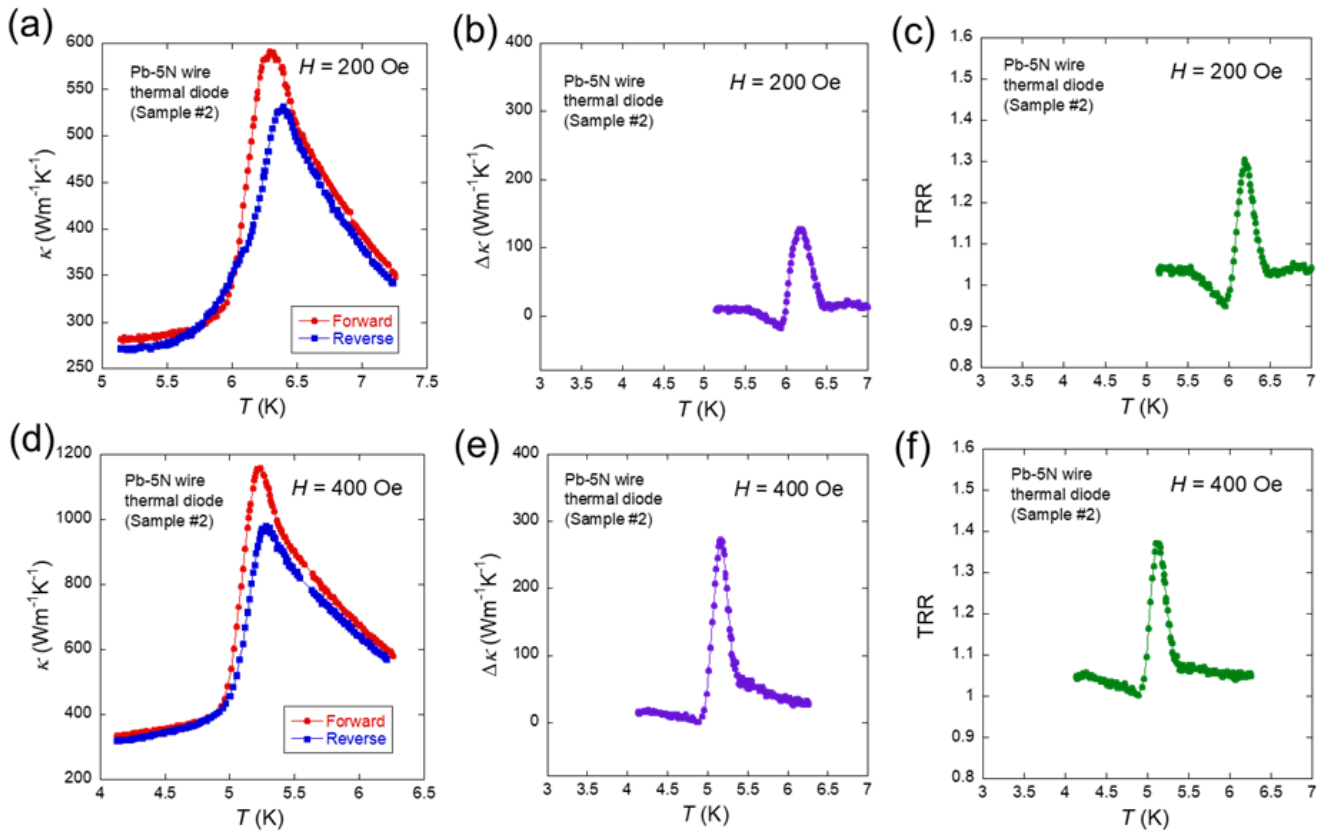

Figure S1. (a–c) Temperature dependence of $\kappa$, $\Delta\kappa$, and thermal rectification ratio (TRR) for the Pb-5N wire thermal diode (sample #2) at $H$ = 200 Oe. (d–f) Temperature dependence of $\kappa$, $\Delta\kappa$, and TRR for the Pb-5N wire thermal diode (sample #2) at $H$ = 400 Oe.

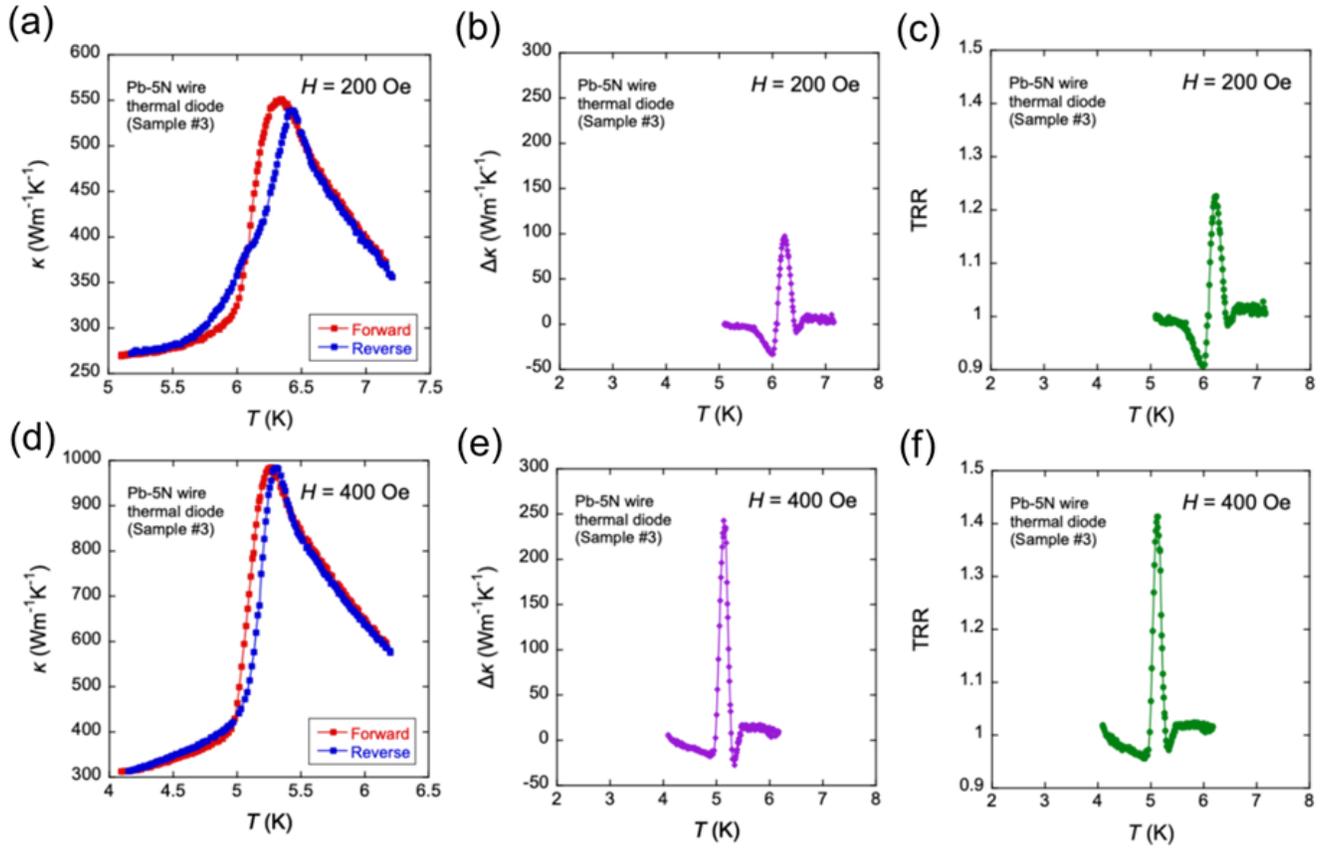

Figure S2. (a–c) Temperature dependence of $\kappa$, $\Delta\kappa$, and TRR for the Pb-5N wire thermal diode (sample #3) at $H = 200$ Oe. (d–f) Temperature dependence of $\kappa$, $\Delta\kappa$, and TRR for the Pb-5N wire thermal diode (sample #3) at $H = 400$ Oe.

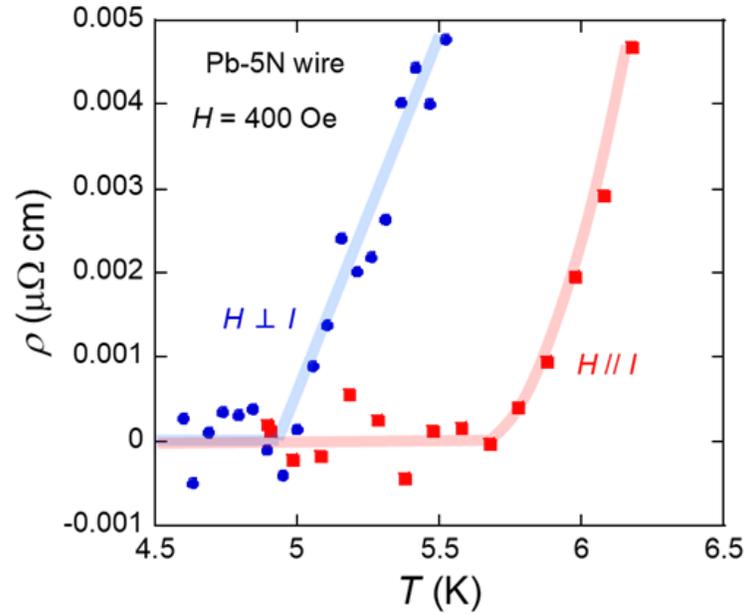

Figure S3. $\rho$-$T$ close to the zero-resistivity state for the Pb-5N wire (resistivity-sample #2) with $H \perp I$ and $H // I$: $H = 400$ Oe. The lines are eye-guide.

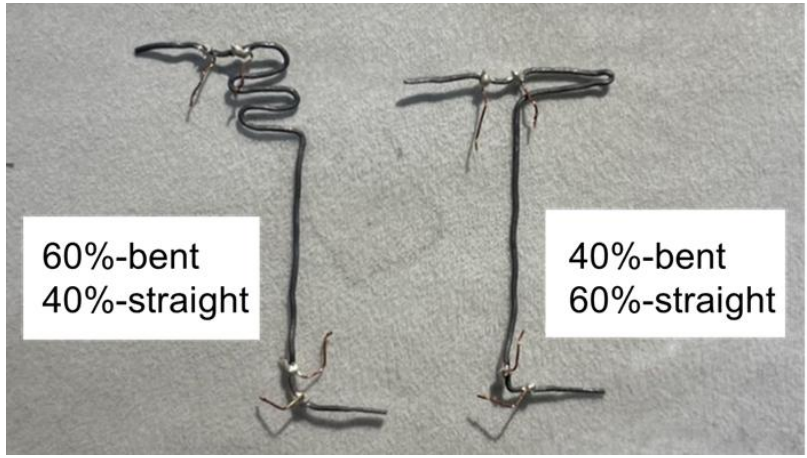

Figure S4. Photo of the 60%-bent and 40%-bent samples.

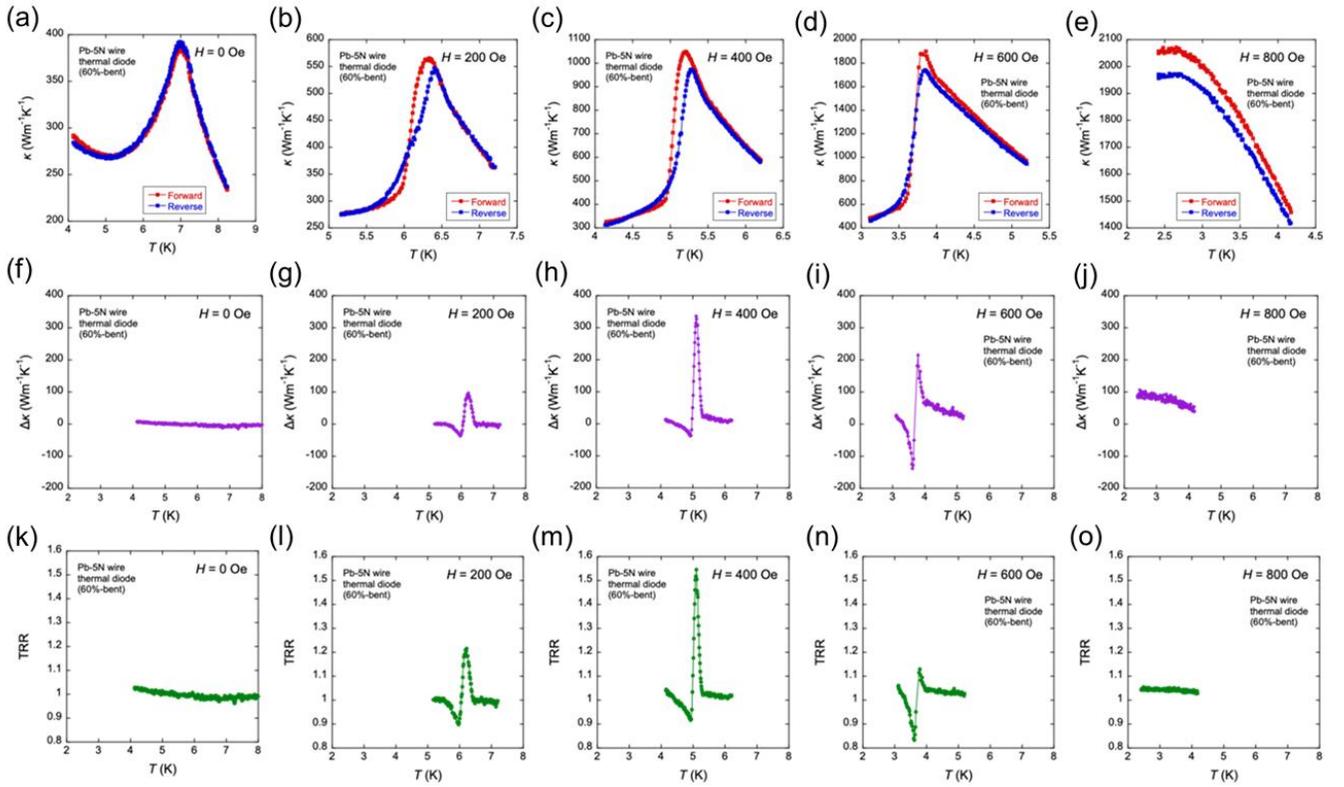

Fig. S5. (a–e) $\kappa$-$T$ of the 60%-bent Pb-5N wire thermal diode measured along the forward and reverse directions at $H =$ (a) 0, (b) 200, (c) 400, (d) 600, and (e) 800 Oe. (f–j) $\Delta\kappa$-$T$ at $H =$ (f) 0, (g) 200, (h) 400, (i) 600, and (j) 800 Oe. (k–o) TRR-$T$ at $H =$ (k) 0, (l) 200, (m) 400, (n) 600, and (o) 800 Oe.

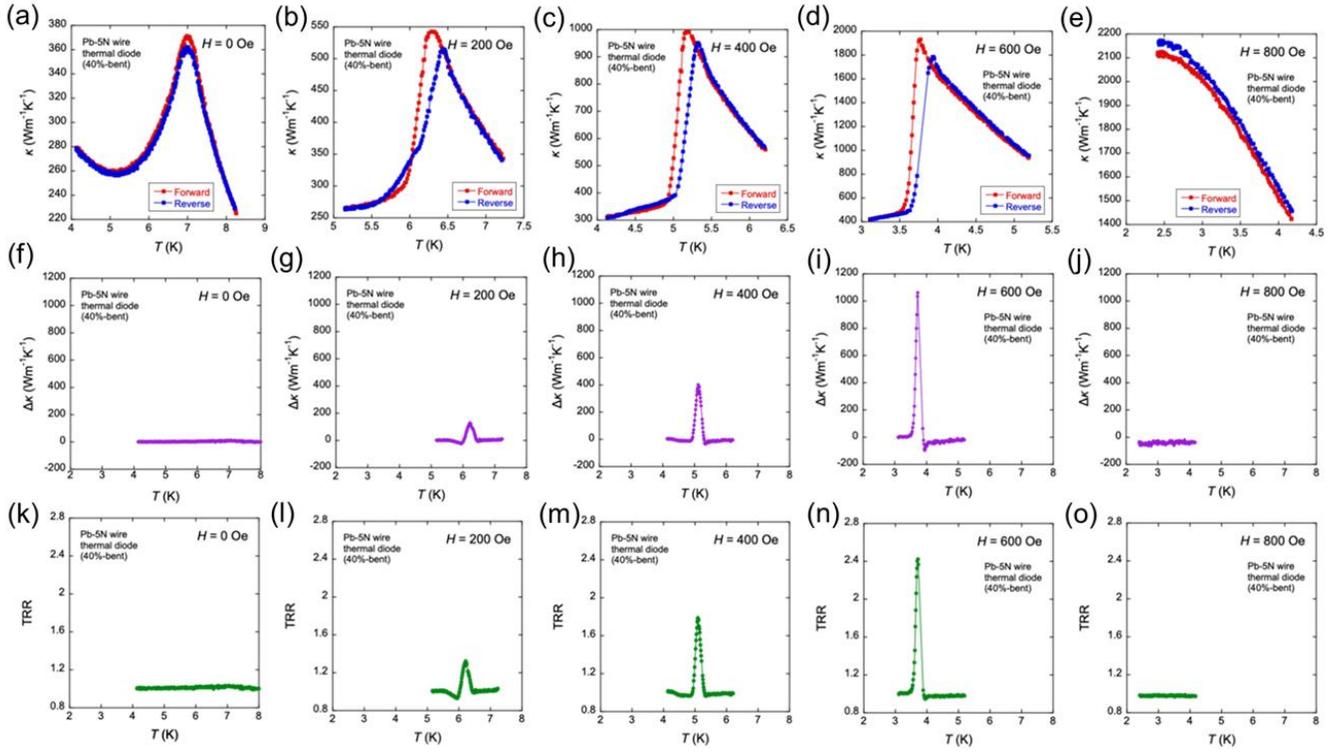

Fig. S6. (a–e) $\kappa$-$T$ of the 40%-bent Pb-5N wire thermal diode measured along the forward and reverse directions at $H =$ (a) 0, (b) 200, (c) 400, (d) 600, and (e) 800 Oe. (f–j) $\Delta\kappa$-$T$ at $H =$ (f) 0, (g) 200, (h) 400, (i) 600, and (j) 800 Oe. (k–o) TRR-$T$ at $H =$ (k) 0, (l) 200, (m) 400, (n) 600, and (o) 800 Oe.